\documentclass[a4paper,11pt]{article}
\usepackage{pos}
\usepackage{float}
\usepackage[symbol]{footmisc}
\usepackage{euscript}
\usepackage{natbib}
\usepackage{doi}
\usepackage{hyperref}
\usepackage{fancybox}
\usepackage{enumerate}
\usepackage{widetext}
\usepackage{subfigure}
\usepackage[compat=1.0.0]{tikz-feynman}

\newcommand{\Meu}{\EuScript{M}}

\def\st{\hbox{}} %% hbox is lower
\usepackage[figuresright]{rotating}
\renewcommand{\thefootnote}{\fnsymbol{footnote}}

\title{Precision multiphoton MC KKMCee for lepton and quark pair production at lepton colliders: Update}
 \ShortTitle{Precision multiphoton MC KKMCee ...}
\centerline{BU-HEPP-26-03, 11/2026}
\author*[a]{B.F.L. Ward}
\author[b]{S. Jadach\footnote[2]{Deceased.}} 
\author[b]{Z. A. Was}
\author[c]{S.A. Yost}
\author[d]{A. Siodmok}

\affiliation[a]{Department of Physics and Astronomy, Baylor University,\\
  Waco, TX, USA\\
and\\
RISC, Johannes Kepler University,\\
Linz, Austria}

\affiliation[b]{Institute of Nuclear Physics,\\
Krakow, PL}

\affiliation[c]{Department of Physics, The Citadel,\\
  Charleston, SC, USA}

\affiliation[d]{Institute of Applied Computer Science, Jagiellonian University,\\
Krakow, PL}

\emailAdd{bfl\_ward@baylor.edu}
\emailAdd{zbigniew.was@ifj.edu.pl}
\emailAdd{yosts1@citadel.edu}
\emailAdd{a.siodmok@cern.ch}

\abstract{We present an overview and update for the {\tt KKMCee 5.00.2} Monte Carlo event generator for lepton and quark pair production in high-energy electron-positron annihilation processes. We note that it is still the most sophisticated event generator for such processes. Its entire source code is rewritten in the modern C++ language and it reproduces all features of the older KKMC code in Fortran 77. We discuss improvements both in the MC algorithm and in its various interfaces, such as those to parton showers and detector simulation.}

\FullConference{43rd International Conference on High Energy Physics (ICHEP2026)\\
30 July to 5 August , 2026\\
Natal, Brazil\\}

\begin{document}
\maketitle

\section{Introduction}
KKMC~\cite{Jadach:1999vf-sh,Arbuzov:2020coe} was developed for precision fermion pair production in $e^+e^-$ collisions, including exponentiated multi-photon effects:
$e^+e^- \rightarrow f \bar{f}+ n\gamma,\; f = \mu,\tau, \nu, u, d, s, c, b,\;\; n = 0, 1, 2, \ldots, \infty,$
including exact $\cal{O}(\alpha), \cal{O}(\alpha^2 L)$ initial and final state photon emission and initial-final
interference, with amplitude-level exponentiation (“CEEX”) inspired by YFS soft photon exponentiation.\par
Currently, $\cal{O}(\alpha)$ EW corrections are included via an independent DIZET v. 6.45 module using a pretabulation approach. Tau decays are carried by a TAUOLA~\cite{Davidson:2010rw} module and collision energies up to 1 TeV are supported with a LEP2 precision  tag of 0.2\%. The most recent release, KKMCee 5.00.2~\cite{Jadach:2022vf}, will be featured in what follows. A significant upgrade from previous versions, it is the first C++ release.\par
In our remaining discussion here, we first present in the next Section a brief review of the CEEX~\cite{Jadach:2000ir} representation of amplitude-based resummation realized in KKMCee before turning to an update  of the elements of KKMCee v. 5.00.2, the new C++ version, in Section 3. Section 4 contains our summary remarks.\par
\section{Review of CEEX YFS Exact Amplitude-Based Resummation}
The basic CEEX~\cite{Jadach:2000ir}  YFS resummation for the prototypical processes $e^+e^-\rightarrow f\bar{f}+n\gamma, \; f = \ell, q, \; \ell=e,\mu,\tau,\nu_e,\nu_\mu,\nu_\tau, \; q = u,d,s,c,b,t$ is given by
\begin{equation}
\sigma =\frac{1}{\text{flux}}\sum_{n=0}^{\infty}\int d\text{LIPS}_{n+2}\; \rho_A^{(n)}(\{p\},\{k\}),
\label{eqn-yfsmth-1}
\end{equation}
where $\text{LIPS}_{n+2}$ denotes Lorentz-invariant phase space for $n+2$ particles, $A=\text{CEEX},\;\text{EEX}$, the incoming and outgoing fermion momenta are abbreviated as 
$\{p\}$, and the $n$ photon momenta are denoted by $\{k\}$. We have (See Refs.~\cite{Jadach:2000ir,Jadach:1999vf}.)
\begin{equation}
\rho_{\text{CEEX}}^{(n)}(\{p\},\{k\})=\frac{1}{n!}e^{Y(\Omega;\{p\})}\bar{\Theta}(\Omega)\frac{1}{4}\sum_{\text{helicities}\;{\{\lambda\},\{\mu\}}}
\left|\Meu\left(\st^{\{p\}}_{\{\lambda\}}\st^{\{k\}}_{\{\mu\}}\right)\right|^2.
\label{eqn-yfsmth-2}
\end{equation}
(The corresponding formula for the $A=\text{EEX}$ case is given in Refs.~\cite{Jadach:2000ir,Jadach:1999vf}.) $Y(\Omega;\{p\})$ is the YFS infrared exponent.
The region $\Omega$ specifies the respective infrared integration limits and its characteristic function
$\Theta(\Omega,k)$ for a photon of energy $k$, with $\bar\Theta(\Omega;k)=1-\Theta(\Omega,k)$ and $$\bar\Theta(\Omega)=\prod_{i=1}^{n}\bar\Theta(\Omega,k_i).$$
By definition, $\Theta(\Omega,k)=1$  for $k\in\Omega$ 
and $\Theta(\Omega,k)=0$ for $k\not\in \Omega$.
We note that, for $\Omega$ defined with the condition $k^0<E_{\min},$ the YFS infrared exponent reads
%//////////////////////////////////////////////////
\begin{equation}
  \label{eq:YFS-ffactor}
  \begin{split}
   &Y(\Omega;p_1,...,p_4)
  =   Q_e^2   Y_\Omega(p_1,p_2)  +Q_f^2   Y_\Omega(p_3,p_4)\\
&\qquad\qquad
     +Q_e Q_f Y_\Omega(p_1,p_3)  +Q_e Q_f Y_\Omega(p_2,p_4) 
     -Q_e Q_f Y_\Omega(p_1,p_4)  -Q_e Q_f Y_\Omega(p_2,p_3),
  \end{split}
\end{equation}
where 
\begin{equation}
\label{eq:form-factor}
Y_\Omega(p,q) 
     \equiv  2 \alpha \tilde{B}(\Omega,p,q)   +2 \alpha \Re B(p,q) 
\end{equation}
is a sum of the real infrared contribution determined by the real emission infrared function $\tilde{B}$ and the virtual infrared contribution determined by the virtual infrared function $B$. The latter infrared functions and the CEEX amplitudes ~$\{\Meu\}$ are defined in Refs.~\cite{Jadach:2000ir,Jadach:1999vf,Jadach:2013aha}. 
The fundamental idea of YFS was to isolate and resum to all orders in $\alpha$ the
infrared singularities so that these singularities are canceled to all such
orders between real and virtual corrections. With CEEX we gain more efficient IFI,
resonance treatment, spin effects including unstable final fermions, ... .\par
\section{Elements of KKMCee 5.00.2 -- The New C++ Version}
There are now two branches of KKMC. One is KKMCee for lepton colliders which is a direct descendant of the original KKMC~\cite{Jadach:1999vf-sh}. The other is KKMChh~\cite{Jadach:2016kh,kkmchh1-sh} for hadron colliders applies the same YFS resummation physics to quark collisions. It has been used in LHC studies but it is not yet publicly released.\par
KKMCee 5.00.2, the C++ version, is publicly available at GitHub: https://github.com/Krakow\newline
HEPSoft/KKMCee. It is documented in Ref.~\cite{Jadach:2022vf}. We note that external libraries such as TAUOLA~\cite{Davidson:2010rw} and DIZET~\cite{dizet:1989} remain in FORTRAN. See Ref.~\cite{dizet642} for documentation of the upgrade of KKMC EW libraries to DIZET 6.45, where DIZET is now realized via pretabulation.\par
We note that in KKMCee 5.00.2, a Les Houches interface facilitates quark pair hadronization. In addition, models of beam energy spread are included for FCC and ILC/CLIC -- the user may include their own. And, the new 5d-semi-analytical tool KKeeFoam~\cite{Jadach:2022vf} is included for cross-checks with analytical resummation.\par
A key improvement in the new C++ version of KKMCee is in event generation. Specifically, FOAM is a general purpose adaptive MC program able to integrate an  arbitrary user provided multi-dimensional distribution. By initial
exploration, an appropriately dense grid of cells is created: peaks, contours, etc. -- enhanced by KKMCee mappings.
In KKMCee, the following are all now generated by FOAM: the ISR energy fraction $v$, beamstrahlung parameters $z_1$ and $z_2\; (z = E/\bar{E})$, the type of final fermion $f = \mu, \tau, \nu_e, \nu_{\mu}, \nu_{\tau}, u, d, s, c, b$,  and the angle $\theta_f$ of the final fermions with respect to the beams. As shown in Ref.~\cite{Jadach:2022vf}, a consequence is that the weight distributions for both s-channel and t-channel involved processes are very much improved. In addition, in the latter reference, the standard benchmarks for 2f production were shown to be reproduced.\par
\begin{figure}[t]
\begin{center}
\setlength{\unitlength}{1in}
\begin{picture}(6.0,3.0)(0,0)
%\put(1.0,0.2){\includegraphics[width=4in]{fig-csfrme-1.pdf}}
\put(.4,0){\includegraphics[width=5in]{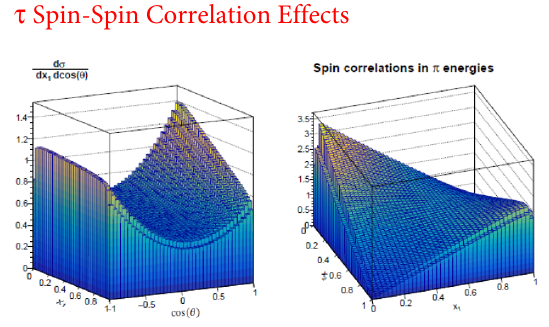}}
%\put(3,0.2){\includegraphics[width=3in]{muon-pt-ratio-6-18-18.pdf}}
\end{picture}
\end{center}
\vspace{-5mm}
\caption{KKMCee predictions for the $\tau$ spin correlation effects. Each $\tau$decays to a single $\pi$ meson: $\tau \rightarrow \pi \nu_{\tau}$. This agrees with Fig.1 in Ref.~\cite{Eberhard:1989ve}.}
\label{fig1}
\end{figure}

Specifically, as an example, we show in Fig.~\ref{fig1} that the LEP results in Ref.~\cite{Eberhard:1989ve} on tau-tau spin correlation effects are reproduced by KKMCee.
\par
\begin{figure}[h!]
\begin{center}
\setlength{\unitlength}{1in}
\begin{picture}(6.0,4.0)(0,0)
%\put(1.0,0.2){\includegraphics[width=4in]{fig-csfrme-1.pdf}}
\put(.9,.2){\includegraphics[width=4in]{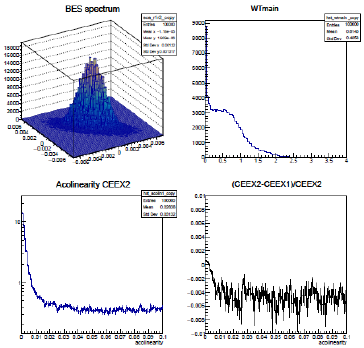}}
%\put(3,0.2){\includegraphics[width=3in]{muon-pt-ratio-6-18-18.pdf}}
\end{picture}
\end{center}
\vspace{-10mm}
\caption{KKMCee BES distribution for 100k events using the new demo program MainBES in Ref.~\cite{sj-bes:2023}. }
\label{fig2}
\end{figure}

\newpage
Finally, we call attention to the FOAM beam energy spread (BES) distributions now available in KKMCee. KKMCee supports bremsstrahlung beam spread (BST) via variables $z_i = E_i/\bar{E}_i$
generated using a CIRCE1~\cite{circe:1996} parametrization, which has sharp peaks at $z_i = 1$, which is now implemented in FOAM. An upgrade to CIRCE2~\cite{whizard-2026} is anticipated. KKMCee now also includes Gaussian beam energy spread (BES)
and FOAM allows the user to provide a combined BES + BST distribution or other custom distribution. We
illustrate this here in Fig.~\ref{fig2}, via a new demo main program MainBES for KKMCee by 
S. Jadach~\cite{sj-bes:2023}. From the 100k events in the figure, using
\par
$$\left(\frac{CEEX2 - CEEX1}{CEEX2}\right)\frac{2\alpha}{\pi}\ln\frac{s}{m_e^2},$$
we get the (crude0 estimated QED uncertainty as $0.02\%$ for accol. below 5 mrads and $0.05\%$ for higher values.\par
\section{Summary Remarks}
We see that the C++ version provides a flexible basis for
future development of KKMCee. Possibilities~\cite{yost-ichep2024} are as follows:
\begin{itemize}
\item Adding CEEX ${\cal O}(\alpha^nL^n), \; n=3,$ corrections which are already in its EEX realization.
\item Forced visible photon at the generator level in neutrino pair channels.
\item Automated CEEX matrix element ($\sim$ 4000 lines of code) construction for porting to other processes such as
$HZ$ production and decay.
\item An  ${\cal O}(\alpha^2)$ EW library. But note that the 1-loop EW corrections to $e^+e^-\rightarrow f \bar{f}\gamma$ are as yet unavailable fully differentially.
\item A more efficient algorithm for certain corners of phase space, such as two hard photons.
\item Integrating the Bhabha process into KKMCee -- essential for FCCee lumi at $10^{-6}$ precision. The latter comment may also apply to $e^+e^-\rightarrow \gamma\gamma$.
\end{itemize}
\section*{Acknowledgments}
Two of us (BFLW and ZW) thank Prof. U. Wiedemann for the support and kind hospitality of the CERN TH Department. This work has been supported by the National Science Centre (NCN), Poland under the grant No. 2023/50/A/ST2/00224. S.A.Yost was supported in part by a grant from the Citadel Foundation.\par
%\baselineskip=10pt
%\begin{thebibliography}{99}
%\bibitem{cern-courier-2023}W. Placzek et al.
\bibliography{Tauola_interface_design}{}
\bibliographystyle{utphys_spires}
%\bibliographystyle{unsrtnat}
%\bibliographystyle{abbrvnat}
%\end{thebibliography}

\end{document}